\documentclass{article}
\usepackage{amsmath,amssymb}
\usepackage{epsfig}

\newtheorem{theorem}{Theorem}
\newtheorem{lemma}{Lemma}

\newtheorem{problem}{Open Problem}
\newtheorem{conjecture}{Conjecture}

\def \lket {|}
\def \rket {\rangle}
\def \lbra {\langle}

\def \A{{\cal A}}
\def \Q{{\cal Q}}
\def\adeg{\widetilde{deg}}
\def\Adv{adv}
\def\bbbc{\mathbb{C}}
\def\bbbn{\mathbb{N}}
\def\bbbr{\mathbb{R}}
\def\N{{\cal N}}

\newcommand{\ket}[1]{\lket #1\rket}

\newcommand{\qed}{\nobreak \ifvmode \relax \else
      \ifdim\lastskip<1.5em \hskip-\lastskip
      \hskip1.5em plus0em minus0.5em \fi \nobreak
      \vrule height0.75em width0.5em depth0.25em\fi}
\newcommand\comment[1]{}

\begin{document}
\title{Understanding Quantum Algorithms via Query Complexity}

\author{Andris Ambainis\thanks{Faculty of Computing,
University of Latvia,
Raina bulv\=aris 19, Riga, LV-1586, Latvia, {\tt ambainis@lu.lv}}}

\date{}

\maketitle

\begin{abstract}
Query complexity is a model of computation in which we have to compute a function $f(x_1, \ldots, x_N)$ of 
variables $x_i$ which can be accessed via queries. The complexity of an algorithm is measured by the number of queries that
it makes.
Query complexity is widely used for studying quantum algorithms, for two reasons.
First, it includes many of the known quantum algorithms (including Grover's quantum search and a key subroutine of Shor's factoring algorithm). Second, one can prove lower bounds on the query complexity, bounding the possible quantum advantage. % (both for specific functions $f(x_1, \ldots, x_N)$ and for general classes of functions). 
In the last few years, there have been major advances on several longstanding problems in the query complexity. In this talk, we survey these results and related work, including:
\begin{itemize}
\item
the biggest quantum-vs-classical gap for partial functions (a problem solvable with 1 query quantumly but requiring $\Omega(\sqrt{N})$ queries classically);
\item
the biggest quantum-vs-determistic and quantum-vs-probabilistic gaps for total functions
(for example, a problem solvable with $M$ queries quantumly but requiring $\tilde{\Omega}(M^{2.5})$ queries probabilistically);
\item
the biggest probabilistic-vs-deterministic gap for total functions
(a problem solvable with $M$ queries probabilistically but requiring $\tilde{\Omega}(M^{2})$ queries deterministically);
\item
the bounds on the gap that can be achieved for subclasses of functions (for example, symmetric functions);
\item
the connections between query algorithms and approximations by low-degree polynomials.
\end{itemize}

\end{abstract}

\section{Introduction}
\label{sec:intro}

Quantum computers open new possibilities for computing, by being able to solve problems that are considered intractable classically. The most famous example is factoring large numbers which is thought to require $\Omega(2^{n^c})$ time classically but is efficiently solvable by a quantum computer, due to Shor's quantum algorithm \cite{Shor}. Another example is simulating quantum physical systems which is thought to require $\Omega(2^{n})$ time classically but is also solvable in polynomial time quantumly \cite{Cirac,Simulation}.

This naturally leads to a question: how large is the advantage of quantum computers? Can we put limits on it?

In the Turing machine model, we have $BQTIME(f(n)) \subseteq \cup_c TIME(2^{c f(n)})$ where $TIME$ and $BQTIME$ denote the classes of problems that are solvable by deterministic or quantum Turing machines within the respective time bound. However, it is very difficult to prove unconditional separations between complexity classes and we cannot even show that $BQTIME(f(n))$ is larger than $TIME(f(n))$. 

For this reason, the power of quantum computers is often studied in the query model  (also known as the decision tree model \cite{BW}). In this model, we have to compute a function $f(x_1, \ldots, x_N)$ of an input $(x_1, \ldots, x_N)$, with $x_i$ accessible via queries to a black box that, given $i$, outputs $x_i$. The complexity is measured by the number of queries that an algorithm makes.

The query model is very interesting in the quantum case because it captures most of the known quantum algorithms. Some of the problems that can be described in it are:

{\bf Search.} Given black box access to $x_1, \ldots, x_N\in\{0, 1\}$, determine whether there exists $i: x_i=1$ (or find such $i$).

Search requires $N$ queries classically but can be solved with $O(\sqrt{N})$ queries quantumly \cite{Grover}. It can be viewed as a black box model for a generic exhaustive search problem where one has to check $N$ possibilities (without any information which of those $N$ possibilities are more likely) and implies quantum speedups for a variety of problems (for example, a quadratic quantum speedup over the best probabilistic algorithm for 3-SAT \cite{SIGACT}).

{\bf Period-finding.} Given black box access to $x_1, \ldots, x_N\in [M]$, determine the smallest $r$ such that $x_i=x_{i+r}$ for all $i$ (and $x_i\neq x_{i+q}$ for all $i$ and $q<r$), under a promise that such $r$ exists and is smaller than $c\sqrt{N}$ for some $c>0$.

Period-finding is solvable with $O(1)$ queries quantumly and requires $\Omega(\frac{N^{1/4}}{\log N})$ queries classically \cite{Shor, Chakraborty}. 
It is at the heart of Shor's factoring algorithm \cite{Shor} which consists of a classical reduction from factoring to period-finding and a quantum algorithm for period-finding.

{\bf Element distinctness.} Given black box access to $x_1, \ldots, x_N\in [M]$, determine if there are $i, j: i\neq j$ such that $x_i = x_j$.

Element distinctness requires $N$ queries classically and $\Theta(N^{2/3})$ queries quantumly \cite{A04,AS}. It is related to black box models of algorithms for breaking collision-resistant hash functions (an important cryptographic primitive). The quantum algorithm for element disticntness is also useful as a subroutine for other quantum algorithms, from checking matrix products \cite{BS} to solving typical instances of subset sum (which is also important for cryptography) \cite{Subset}.

Many other quantum query algorithms are known, as can be seen from Quantum Algorithms Zoo (a website collecting information about all quantum algorithms \cite{Zoo}). % lists xx quantum algorithms and yy\% of them can be either described in the query model or %have important subroutines that can be described in the query model. 
From a complexity-theoretic perspective, the query model is very interesting because it allows to prove lower bounds on quantum algorithms and it is often possible to characterize the quantum advantage within a big-O factor. 

The current survey is focused on characterizing the maximum possible quantum advantage in the query model, for different types of computational tasks. 
Let $Q(f)$ and $R(f)$ denote the number of queries for the best quantum and randomized algorithm, respectively. For partial Boolean functions, we describe a gap of $Q(f)=1$ vs $R(f)=\Omega(\sqrt{N}/\log N)$ \cite{AA}. For total functions, the biggest known gap is much smaller: $R(f)=\tilde{\Omega}(Q^{2.5}(f))$ \cite{ABK}. Imposing symmetry constraints on $f$ also decreases the maximum possible gap.

As a side result, this research has lead to new results on classical query algorithms. This includes solutions to two well known problems in the classical query complexity which had been open for about 30 years (such as determining the maximum gap between randomized and deterministic query complexities \cite{SW,AB+}). We describe those developments, as well.

\section{Computational Models}

\subsection{Deterministic, randomized and quantum query algorithms}

We now formally define the models of query complexity that we use. We consider computing a function $f(x_1, \ldots, x_N)$ of variables $x_i$.  By default, we assume that the variables $x_i$ are $\{0, 1\}$-valued. (If we consider $x_i$ with values in a larger set, this is explicitly indicated.) The function $f(x_1, \ldots, x_N)$ can be either a total function (defined on the entire $\{0, 1\}^N$) or a partial function (defined on a subset of $\{0, 1\}^N$).

\begin{figure}[ht!]
\centering
\hspace{0in}
\epsfxsize=2in
\epsfbox{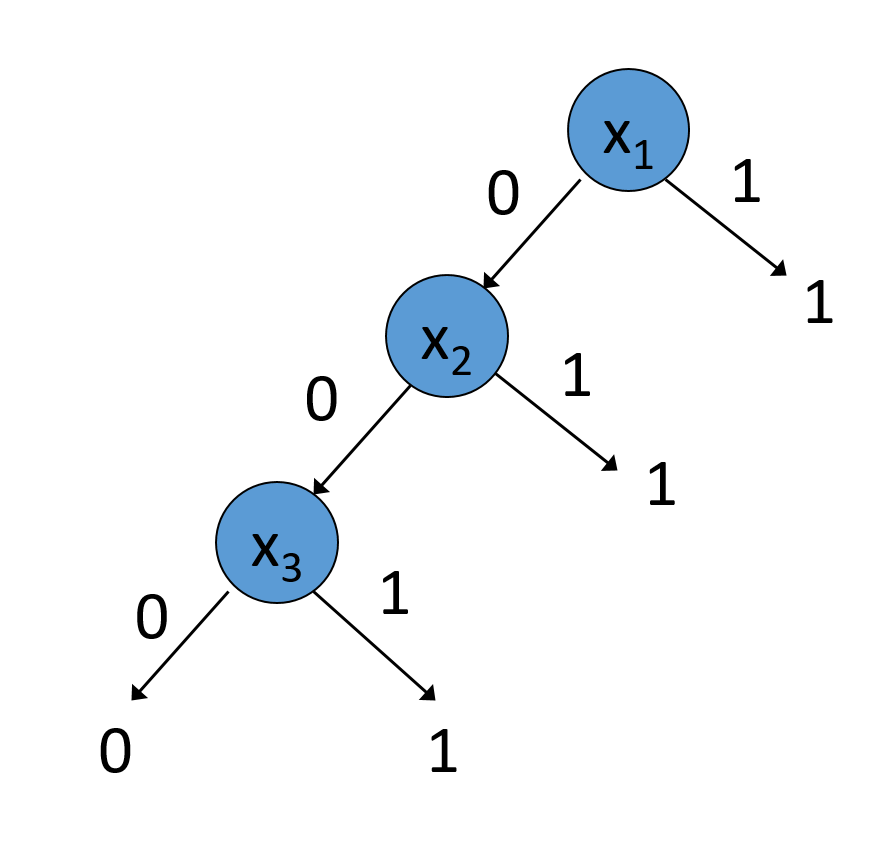}
\caption{Example of a decision tree}
\label{fig:dec-tree}
\end{figure}

{\bf Deterministic algorithms.} Deterministic query algorithms are often called {\em decision trees}, because they can be described by a tree (as in figure \ref{fig:dec-tree}).  At each node of this tree, we have the name of a variable that is asked if the algorithm gets to this node. Depending on the outcome of the query, the algorithm proceeds to the $x_i=0$ child or to the $x_i=1$ child of the node. If the algorithm gets to a leaf of the tree, it outputs the value of the function listed at this leaf.

The complexity of an algorithm ${\cal A}$ is  the maximum number of queries that it can make. Deterministic query complexity $D(f)$ is the smallest complexity of a deterministic ${\cal A}$ which outputs $f(x_1, \ldots, x_N)$ if the queries are answered according to $(x_1, \ldots, x_N)$, whenever $f(x_1, \ldots, x_N)$ is defined.

{\bf Randomized algorithms.} In a randomized query algorithm, the algorithm may choose the variable $x_i$ for the next query randomly from some probability distribution. %For any such algorithm, it is possible to make all random choices at the beginning of algorithm. Therefore, a randomized query algorithm can be viewed as a collection of deterministic decision trees $T_i$ with probabilities $p_i$. At the beginning one of $T_i$'s is chosen. After that, the computation proceeds according to $T_i$.

Randomized algorithms are usually studied either in the zero-error setting or in the bounded error setting. In the zero-error setting, the algorithm is required to output $f(x_1, \ldots, x_N)$ with probability at least 1/2 and may output "don't know" otherwise but must not output a value that is different from $f(x_1, \ldots, x_N)$. In the bounded-error setting, algorithm is required to output $f(x_1, \ldots, x_N)$ with probability at least 2/3 and may output anything otherwise.
In both cases, the requirement has to be satisfied 
for every $(x_1, \ldots, x_N)$ for which $f(x_1, \ldots, x_N)$ is defined.

The complexity of an algorithm ${\cal A}$ is measured by the largest number of queries that is made by ${\cal A}$, for the worst choice of $(x_1, \ldots, x_N)$ and the worst random choices of ${\cal A}$. $R_0(f)$ and $R_2(f)$ are the smallest complexities of a zero-error randomized and a bounded error randomized algorithm for $f$, respectively. (Alternatively, one can define randomized query complexity via the expected number of queries for the worst case $(x_1, \ldots, x_N)$
but this changes the complexities $R_0$ and $R_2$ by at most a constant factor.)

{\bf Quantum algorithms.} Unlike in the probabilistic case, different branches of a quantum algorithm can recombine at a later stage. For this reason, a quantum query algorithm cannot be described by a tree.

\begin{figure}[h!]
\centering
\hspace{0in}
\epsfxsize=5in
\epsfbox{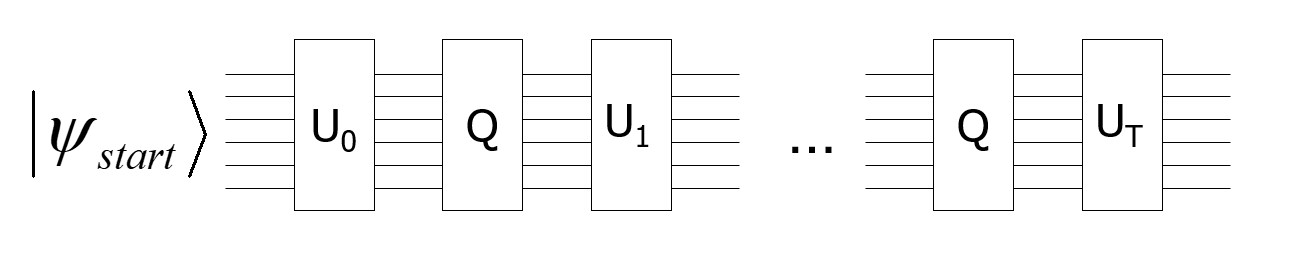}
\caption{Structure of a quantum query algorithm}
\label{fig:quantum}
\end{figure}

Instead, a quantum query algorithm is defined by an initial state $\ket{\psi_{start}}$ and transformations $U_0, Q, U_1, \ldots, Q, U_T$. The initial state $\ket{\psi_{start}}$ and transformations $U_i$ are independent of $x_1, \ldots, x_N$. $Q$ are the queries - transformations of a fixed form that depend on $x_i$'s. The algorithm consists of performing $U_0, Q, U_1, \ldots, Q, U_T$ on $\ket{\psi_{start}}$ and measuring the result (as shown in Figure \ref{fig:quantum}). The algorithm computes $f$ if, for every $(x_1, \ldots, x_N)$ for which $f(x_1, \ldots, x_N)$ is defined, this measurement produces $f(x_1, \ldots, x_N)$. %, either always or with a high probability (depending on the model of error that is chosen).

To define the model more precisely, we must define the notions of a quantum state, a transformation, and a measurement. (For more details on these notions, we refer the reader to the book \cite{NC}.)
The state space of a quantum algorithm is a complex vector space of dimension $d$ (where $d$ can be choosen by the designer of the algorithm). Let $\ket{1}, \ldots, \ket{d}$ be an orthonormal basis for this vector space. A quantum state is a vector 
\[ \ket{\psi} = \alpha_1 \ket{1}+ \ldots + \alpha_d \ket{d} = 
\left( \begin{array}{c} \alpha_1 \\ \alpha_2 \\ \ldots \\ \alpha_d \end{array} \right) \]
of unit length (i.e. satisfying $\sum_i |\alpha_i|^2 =1$). A unitary transformation is a linear transformation on $\ket{\psi}$ that preserves the length of $\ket{\psi}$. The principles of quantum mechanics allow to perform any unitary $U$ on a quantum state.

A measurement is the way of obtaining information from a quantum state. Measuring a state $\ket{\psi}$ with respect to $\ket{1}, \ldots, \ket{d}$ %\footnote{Other measurements are also possible but will not be important for the purposes of this survey.} 
yields the result $i$ with probability $|\alpha_i|^2$.

To define a quantum query algorithm, we allow the starting state $\ket{\psi_{start}}$ to be an arbitrary quantum state. $U_i$'s can be arbitrary unitary transformations that do not depend on $x_1, \ldots, x_N$. $Q$ is the query transformation, defined in a following way\footnote{Since most of this survey considers functions $f(x_1, \ldots, x_N)$ of variables $x_i\in\{0, 1\}$, we only give the definition of a query for this case.}. We rename the basis states from $\ket{1}, \ldots, \ket{d}$ to 
$\ket{i, j}$ with $i\in\{0, 1, \ldots, N\}$ and $j\in [d_i]$ for some $d_i$ and define
\[ Q\ket{0, j} = \ket{0, j} \mbox{~for all~} j ,\]
\[ Q\ket{i, j} = \begin{cases} 
\ket{i, j} & \text{if $x_i = 0$} \\
-\ket{i, j} & \text{if $x_i = 1$} 
\end{cases} .\]
It can be argued that this is a natural quantum conterpart of a probabilistic query in which we choose $i$ according to a probability distribution and get the corresponding $x_i$.

After the last transformation, the state of the algorithm is measured w.r.t. $\ket{1}, \ldots, \ket{d}$ and the result is transformed into the answer of the algorithm according to a predefined rule. (For example, if the answer should be $\{0, 1\}$-valued, we could take the first bit of the measurement result $i$ as the answer.) 

Two most frequently considered types of quantum query algorithms are {\em exact} and {\em bounded error} algorithms. A quantum query algorithm computes $f$ exactly if its answer is always the same as $f(x_1, \ldots, x_N)$, whenever $f$ is defined. A quantum query algorithm $\A$ computes $f$ with bounded error, if for every $(x_1, \ldots, x_N)$, for which $f(x_1, \ldots, x_N)$ is defined, the probability that $\A$ outputs $f(x_1, \ldots, x_N)$ as the answer is at least 2/3. $Q_E(f)$ and $Q_2(f)$ are the smallest numbers of queries in quantum algorithms that compute $f$ exactly and with bounded error, respectively.

\subsection{Quantities that are related to query complexity}

In this section, we define several quantities that provide upper and lower bounds on different query complexities. Using them, we can prove bounds on the maximum gaps between query complexity measures (for example, that $D(f)=O(R_2^3(f))$ \cite{Nisan} and $D(f)=O(Q_2^6(f))$  \cite{Beals} for any total Boolean function $f$).

{\bf Block sensitivity.} For an input $x\in\{0, 1\}^N$ and a subset of variables $S\subseteq [N]$, $x^{(S)}$ is the input obtained from $x$ by changing all $x_i, i\in S$ to opposite values. The block sensitivity $bs(f)$ is the maximum $k$ for which there is an input $x\in\{0, 1\}^N$ and pairwise disjoint subsets $S_1, \ldots, S_k\subseteq [N]$ with $f(x)\neq f(x^{(S_i)})$ for all $i\in [k]$.

Block sensitivity is a lower bound on all the query complexity measures: $D(f)\geq bs(f)$, $R(f) = \Omega(bs(f))$ \cite{Nisan} and $Q(f) = \Omega(\sqrt{bs(f)})$ \cite{Beals}. 
It also provides an upper bound on $D(f)$ for total Boolean functions $f$: $D(f)=O(bs^3(f))$ \cite{Nisan}. Combining these relations yields $D(f)=O(R_2^3(f))$  and $D(f)=O(Q_2^6(f))$ - the
 best upper bounds on the gap between $D(f)$ and $R_2(f)$ or $Q_2(f)$.

{\bf Certificate complexity.} For an input $x\in\{0, 1\}^N$, a certificate is a set $S\subseteq [N]$ with the property that the variables $x_i, i\in S$ determine the value of $f(x)$.
(More precisely, $S\subseteq [N]$ is a certificate on an input $x$ if, for any $y\in\{0, 1\}^N$ such that $x_i=y_i$ for all $i\in [S]$, we have $f(x)=f(y)$.)
$C_x(f)$ is the minimum size $|S|$ of a certificate $S$ on the input $x$. 
The certificate complexity $C(f)$ is the maximum of $C_x(f)$ over all $x\in\{0, 1\}^N$.

Certificate complexity provides a better upper bound on $D(f)$ for total $f$: $D(f)\leq C^2(f)$ \cite{Nisan}. 
If one could show that $Q_2(f)=\Omega(\sqrt{C(f)})$, this would imply $D(f)=O(Q_2^4(f))$, improving the best known relation between $D(f)$ and $Q_2(f)$.

{\bf Randomized certificate complexity \cite{Aaronson2008}.} For an input $x$, $RC_x(f)$ is the minimum number of queries in a bounded-error randomized query algorithm that accepts $x$ and rejects all $y: f(x) \neq f(y)$. The randomized certificate complexity $RC(f)$ is the maximum of $RC_x(f)$ over all $x\in\{0, 1\}^N$.  

Unlike for the standard certificate complexity, it is known that $Q_2(f)=\Omega(\sqrt{RC(f)})$ \cite{Aaronson2008}.
Proving $D(f)=O(RC^2(f))$ for total $f$ (which is not known) would also imply $D(f)=O(Q_2^4(f))$.

{\bf Polynomial degree.} The exact degree, $\deg(f)$, is the degree of the multilinear polynomial $p(x_1, \ldots, x_N)$ which satisfies $f(x_1, \ldots, x_N)=p(x_1, \ldots, x_N)$ for all $(x_1, \ldots, x_N)$. The approximate degree, $\adeg(f)$, is the smallest degree of a multilinear polynomial $p(x_1, \ldots, x_N)\in\{0, 1\}^N$ which satisfies $|f(x_1, \ldots, x_N)-p(x_1, \ldots, x_N)|\leq \frac{1}{3}$ for all $(x_1, \ldots, x_N)\in\{0, 1\}^N$. 

Both of these measures also provide lower bounds on quantum query complexity: $Q_E(f)\geq \frac{\deg(f)}{2}$ and $Q_2(f) = \Omega(\sqrt{\adeg(f)})$ \cite{Beals}.

\section{Maximum quantum-classical gap for partial functions}
\label{sec:partial}

In this section, we consider the question: what is
the maximum possible gap between $Q_2(f)$ and the most general of classical complexities $R_2(f)$, for a partial function $f(x_1, \ldots, x_N)$ if we do not place any constraints on $f$?

As we already mentioned, period finding has $Q_2(f)=O(1)$ and $R_2(f)=\tilde{\Omega}(\sqrt[4]{N})$. In the form defined in section \ref{sec:intro}, period-finding is not a Boolean function (it has variables
$x_i\in[M]$ instead of Boolean variables). While it is possible to define a Boolean version of period-finding with almost the same gap, there is Boolean function with an even bigger gap:

\begin{theorem}
\label{thm:forrelation}
\cite{AA}
There exists $f$ with $Q_2(f)=1$ and $R_2(f)=\Omega(\sqrt{N}/\log N)$.
\end{theorem}

The function $f$ is defined as follows \cite{AA}. We have $N=2^{n+1}$ variables. For technical convenience,  we denote variables $x_0, \ldots, x_{2^{n}-1}, y_0, \ldots, y_{2^n-1}$ and assume that the possible values for variables are $\pm 1$ (instead of 0 and 1). Let $F$ be the $2^n \times 2^n$ matrix (with rows and columns indexed by $a, b\in [0, 2^n-1]$) defined by $F_{a, b} = \frac{1}{2^{n/2}} (-1)^{a\cdot b}$ where $a\cdot b = \sum_i a_i b_i$
is the inner product between $a$ and $b$ interpreted as $n$-bit strings $a_{n-1}\ldots a_0$ and $b_{n-1}\ldots b_0$. (In terms of quantum computing, $F=H^{\otimes n}$ where $H$ is the standard $2\times 2$ Hadamard matrix.) We define
\[ f(x_0, \ldots, y_{2^n-1}) = \begin{cases}
1 & \mbox{~if~} \sum_{a, b} F_{a, b} x_a y_b \geq \frac{3}{5} 2^n \\
0 & \mbox{~if~} \sum_{a, b} F_{a, b} x_a y_b \leq \frac{1}{100} 2^n 
\end{cases}.\]
The thresholds $\frac{3}{5}$ and $\frac{1}{100}$ are chosen so that:
\begin{itemize}
\item
if we choose $x_i\in\{-1, 1\}$ for $i\in\{0, \ldots, 2^n-1\}$ uniformly at random and then choose $y_i=sgn((Fx)_i)$, we get $f=1$ with a high probability;
\item
if we choose both $x_i$ and $y_j$ uniformly at random form $\{-1, 1\}$, we get $f=0$ with a high probability.
\end{itemize}
Thus, by solving $f$, we are effectively distinguishing between $\vec{y}=(y_i)_{i\in[0, 2^n-1]}$ being the vector of signs of $F\vec{x}$ where
$\vec{x}=(x_i)_{i\in[0, 2^n-1]}$ and $\vec{y}$ being independently random.

$Q_2(f)=1$ is shown by a quantum algorithm that generates a quantum state 
\[ \ket{\psi}= \sum_{i=0}^{2^n-1} \left( \frac{x_i}{\sqrt{2^n}} \ket{0, i} + \frac{y_i}{\sqrt{2^n}} \ket{1, i} \right) .\]
This quantum state can be generated by just 1 query. We then apply the transformation $F$ to basis states $\ket{0, i}$,
transforming the state to 
\[ \ket{\psi}= \sum_{i=0}^{2^n-1} \left( \frac{(Fx)_i}{\sqrt{2^n}} \ket{0, i} + \frac{y_i}{\sqrt{2^n}} \ket{1, i} \right) .\]
We then use the SWAP test \cite{Fingerprinting}, a well known test for testing similarity of coefficient vectors of two parts of a quantum state.

The proof of the lower bound, $R_2(f)=\Omega(\sqrt{N}/\log N)$, is quite intricate. We define a corresponding problem (which we call REAL FORRELATION) with real valued variables 
$x_0, \ldots, x_{2^{n}-1}, y_0, \ldots, y_{2^n-1}$ in which we have to distinguish between two cases:
\begin{enumerate}
\item[(a)]
all $x_i$ and $y_i$ are i.i.d. random with Gaussian distribution $\N(0, 1)$;
\item[(b)]
$x_i$'s are i.i.d random with Gaussian distribution $\N(0, 1)$ and $y_i$ are obtained by applying Fourier transform to a vector consisting of $x_i$'s: $y_i=((Fx)_i)$.
\end{enumerate}
In \cite{AA}, we show that any algorithm for FORRELATION implies an algorithm for REAL FORRELATION with a similar complexity. Thus, it suffices to show a classical lower bound on
REAL FORRELATION.

REAL FORRELATION is, in turn, a special case of a more general problem, GAUSSIAN DISTINGUISHING, in which we have to determine whether a set of real-valued variables $x_1, \ldots, x_M$ has a hidden structure. Let
$\vec{v_1}$, $\ldots$, $\vec{v_M}$ be a set of vectors in $\bbbr^d$ for some $d$. We have to distinguish between two cases:
\begin{enumerate}
\item[(a)]
all $x_i$ are i.i.d. random with Gaussian distribution $\N(0, 1)$;
\item[(b)]
$x_1, \ldots, x_M$ are generated by choosing a random $\vec{u}\in \bbbr^d$ (whose entries are i.i.d. $\N(0, 1)$ random variables) and taking $x_i= (\vec{u}, \vec{v_i})$.
\end{enumerate}

The lower bound on REAL FORRELATION is a special case of 
\begin{theorem}
Let $\vec{v_i}$ be such that $|(\vec{v_i}, \vec{v_j})| \leq \epsilon$ for all $i\neq j$. Then,
GAUSSIAN DISTINGUISHING requires $\Omega(\frac{1/\epsilon}{\log (M/\epsilon)})$ queries.
\end{theorem}

In the case of REAL FORRELATION, $M=2^{n+1}$, $d=2^n$, $\vec{v_1}, \ldots, \vec{v}_{2^n}$ are the computational basis
states $\ket{0}$, $\ldots$, $\ket{2^n-1}$ and $\vec{v}_{2^n+1}, \ldots, \vec{v}_{2^{n+1}}$ are 
$F\ket{0}$, $\ldots$, $F\ket{2^n-1}$. Then, $\epsilon=\frac{1}{\sqrt{2^n}}=\frac{1}{\sqrt{N/2}}$, implying a lower bound
of $\Omega(\sqrt{N}/\log N)$ on REAL FORRELATION. This bound is nearly tight, as shown by

\begin{theorem}
\label{thm:qs}
Let $\A$ be a 1-query quantum algorithm. There is a probabilistic algorithm $\A'$ that makes $O(\sqrt{N})$ queries
and, on every input $(x_1, \ldots, x_N)$, outputs an estimate $\tilde{p}$ such that $|p-\tilde{p}|\leq \epsilon$
(where $p$ is the accepting probability of $\A$ on $(x_1, \ldots, x_N)$) with a high probability.
\end{theorem}

The simulation makes use of the connection between quantum algorithms and polynomials:

\begin{lemma}
\label{lem:poly}
\cite{Beals}
Let $\A$ be a quantum algorithm that makes $k$ queries to an input $(x_1, \ldots, x_N)$, $x_i\in\{0, 1\}$. 
The accepting probability of $\A$ can be expressed as 
a polynomial $p(x_1, \ldots, x_N)$ in variables $x_1, \ldots, x_N$ of degree at most $2k$.
\end{lemma}

Since the accepting probability of an algorithm must be between 0 and 1, we have
$0\leq p(x_1, \ldots, x_N)\leq 1$ whenever $x_1, \ldots, x_N\in\{0, 1\}$. Theorem \ref{thm:qs} then follows
from a more general result about estimating bounded polynomials:

\begin{lemma}
\cite{AA}
For every polynomial $p(x_1, \ldots, x_N)$ with $\deg p \leq 2$ and $0\leq p(x_1, \ldots, x_N)\leq 1$ 
for any $x_1, \ldots, x_N\in\{0, 1\}$, there is a probabilistic algorithm $\A'$ that makes $O(\sqrt{N})$ queries
and outputs an estimate $\tilde{p}$ such that 
$|p(x_1, \ldots, x_N)-\tilde{p}|\leq \epsilon$ with a high probability, for every input $(x_1, \ldots, x_N)\in\{0, 1\}^N$.
\end{lemma}

More generally, if we have a bounded polynomial $p(x_1, \ldots, x_N)$ with $\deg p \leq k$, its value can be estimated
with $O(N^{1-1/k})$ queries. Together with lemma \ref{lem:poly}, this implies a probabilistic
simulation of $t$ query quantum algorithms with $O(N^{1-1/2t})$ queries. Unlike for $t=1$, we do not know whether this
is optimal.

\begin{problem}
\label{prob:gap}
Let $t\geq 2$. Is there a partial function $g(x_1, \ldots, x_N)$ with $Q_2(g)=t$ and $R_2(g)=\tilde{\Omega}(N^{1-1/2t})$?
\end{problem}

The problem remains open, if instead of $\tilde{\Omega}(N^{1-1/2t})$ (which matches our upper bound), we ask for a 
weaker lower bound of $\Omega(N^c)$, $c>1/2$ and, even, if instead of a constant $t$, we allow $t=O(\log^c N)$.

\begin{problem}
Is there a partial function $g(x_1, \ldots, x_N)$ with $Q_2(g)=O(\log^c N)$ for some $c$ and $R_2(g)=\Omega(N^d)$ for $d>1/2$?
\end{problem}

%The number of candidate problems is fairly small. 
The well known examples of problems with a large quantum-classical gap
(such as Simon's problem \cite{Simon} or period-finding) typically have $R_2(g)=O(\sqrt{N})$. In \cite{AA}, we give a candidate problem, 
$k$-FOLD FORRELATION for which we conjecture that bounds of Problem \ref{prob:gap} hold. This is, however, the only
candidate problem that we know.

\section{Total functions: pointer function method}
\label{sec:total}

For total functions $f$, the possible gaps between $Q(f)$, $R(f)$ and $D(f)$ are much smaller: all of these
complexity measures are polynomially related. 

It is well known that $D(f)=O(Q_2^6(f))$ \cite{Beals} and $D(f)=O(R_2^3(f))$ \cite{Nisan}. 
For exact/zero error algorithms we know that $D(f)=O(Q_E^3(f))$ \cite{Midrijanis} and $D(f)=O(R_0^2(f))$ \cite{SW}.
The question is: how tight are these bounds?

For a very long time, the best separations were:
\begin{itemize}
\item
Quantum vs. probabilistic/deterministic: $OR(x_1, \ldots, x_N)$ has $Q_2(OR)=O(\sqrt{N})$ due to Grover's 
quantum search algorithm and $R_2(f)=\Omega(N)$. 
\item
Probabilistic vs. deterministic: the binary AND-OR tree function of depth $d$ has $D(f)=2^d$ and 
$R_0(f)=O((\frac{1+\sqrt{33}}{4})^d)$, implying that $R_0(f)=O(D^{0.753...}(f))$ \cite{SW}.
\end{itemize}

Both of these separations were conjectured to be optimal by a substantial part of the respective research community.

For exact quantum query complexity, the best separation was $Q_2(XOR)=N/2$ vs. $D(XOR)=R_2(XOR)=N$ for the $N$-bit XOR function \cite{Beals} until 
2013 when an example with $Q_E(f)=O(R^{0.86...}_2(f))$ was discovered \cite{A13}.

In 2015, major improvements to all of these bounds were achieved, via two new methods. 
The first of them, the {\em pointer function} method was first  invented by G\"o\"os, Pitassi and Watson \cite{GPW} for
solving the communication vs. partition number problem in classical communication complexity.
It was then quickly adapted to separating query complexity measures by Ambainis et al. \cite{A+16}:

\begin{theorem}
\cite{A+16}
\label{thm:amain}
\begin{enumerate}
\item
There exists a total Boolean function $f$ with $Q_2(f)=\tilde{O}(D^{1/4}(f))$.
\item
There exists a total Boolean function $f$ with  $R_0(f)=\tilde{O}(D^{1/2}(f))$.
\item
There exists a total Boolean function $f$ with $R_2(f)=\tilde{O}(R_0^{1/2}(f))$.
\end{enumerate}
\end{theorem}

The first two results provide major improvements over the previosly known results mentioned at the beginning of this section.
The third result is the first ever superlinear gap between $R_0(f)$ and $R_2(f)$ for a total $f$.

We now illustrate the method by describing the simplest function by G\"o\"os, Pitassi and Watson \cite{GPW}
and sketch a proof that it achieves $R_2(f)=\tilde{O}(D^{1/2}(f))$, a slightly weaker result than
the second item above.
Consider $f(x_{ij}, y_{ij}, z_{ij})$, with variables $x_{ij}\in\{0, 1\}, y_{ij}\in [0, N], z_{ij}\in [0, M]$ indexed by $i \in [N], j \in [M]$.
The variables $x_{ij}$ are interpreted as elements of an $N\times M$ table and pairs of variables $(y_{ij}, z_{ij})$ are
interpreted as pointers to entries in this table\footnote{As described, this is a function of variables with a larger set of values but it can be converted into a function with $\{0, 1\}$-valued variables, with complexities changing by at most a logarithmic factor.} .

We define that $f=1$ if the following conditions are satisfied:
\begin{enumerate}
\item
the $N\times M$ table has a unique column $i$ in which all entries $x_{ij}$ are 1;
\item
in this column, there is exactly one $j$ for which $(y_{ij}, z_{ij})\neq (0, 0)$; 
\item
if we start at this $(i, j)$ and repeatedly follow the pointers (that is, consider the sequence $(i_k, j_k)$ defined
by $(i_0, j_0)=(i, j)$ and $(i_k, j_k) = (y_{i_{k-1}j_{k-1}}, z_{i_{k-1}j_{k-1}})$ for $k>0$), then:
\begin{enumerate}
\item
for each $i'\neq i$, there is a unique $k\in [N-1]$ with $i_k=i'$,
\item
$(i_N, j_N)=(0, 0)$,
\item
$x_{i_k j_k}=0$ for all $k\in [N-1]$.
\end{enumerate}
\end{enumerate}

This function $f$ has the following properties:
\begin{enumerate}
\item
$D(f)=NM$: for any deterministic algorithm, an adversary may choose the values for variables so that at least one of $x_{ij}$, $y_{ij}$, $z_{ij}$ needs to be queried for each $ij$.
\item
If $f(x_{ij}, y_{ij}, z_{ij}) = 1$, this can be certified by showing variables $x_{ij}, y_{ij}, z_{ij}$ for $N+M-1$ different $(i, j)$: the all-1 column and the cells $(i_k, j_k)$ in the sequence of pointers. Moreover, there is one and only one way to certify this.
\end{enumerate}

To show a gap between $D(f)$ and $R_2(f)$, it suffices to show that a randomized algorithm can find this certificate faster than a deterministic algorithm. For that, we set $N=M$ and consider the following randomized algorithm (due to Mukhopadhyay and Sanyal \cite{MS}):
\begin{enumerate}
\item
$\Theta(N \log N)$ times repeat:
\begin{enumerate}
\item
Choose a random entry $(i, j)$ of the table in a column that has not been eliminated yet.
\item
While $x_{ij}=0$, $y_{ij}\neq 0$, $z_{ij}\neq 0$ and $i$ is not a column that has been already eliminated:
\begin{itemize}
\item
eliminate column $i$;
\item
set $i=y_{ij}$ and $j=z_{ij}$.
\end{itemize}
\item
If $x_{ij}=0$ but $y_{ij}=0$ or $z_{ij}=0$, eliminate column $i$. 
\end{enumerate}
\item
If all columns are eliminated or more than 100 columns remain, output 0.
\item
Otherwise, test each of remaining columns by checking whether it satisfies the conditions for a certificate. 
\end{enumerate}

If $f=1$, each time when we choose a random entry in a column that is not the all-1 column, 
there is an $\frac{1}{N}$ probability of choosing the entry that is a part of the pointer chain.
This means that, during $\Theta(N \log N)$ repetitions, this happens $\Theta(\log N)$ times. 
Each time, the columns that are after this entry in the pointer chain get eliminated. On
average, half of remaining columns are after the entry that gets chosen. This means that, with a high probability,
after $\Theta(\log N)$ times, only $O(1)$ columns are not eliminated. Then, one can test each of them with $O(N)$ queries.

This basic construction can be modified  
in several ways \cite{A+16}. To separate two models of computation, we should make the certificate for $f=1$
easy to find in one of them but difficult in the other model. (For example, hard to find by zero-error probabilistic 
algorithms but easy to find by bounded error probabilistic algorithms.) For different separations, the modifications include:
\begin{itemize}
\item
Arranging the cells with pointers (in columns that are not the all-1 column) into a binary tree instead of a chain.
\item
Introducing back pointers at the end of the pointer chain or at the leaves of the tree, pointing back to the all-1 column.
\item
Having more than one all-1 column with pointers among the all-1 columns.
\end{itemize}

Besides the three major results in Theorem \ref{thm:amain}, this approach gives better-than-before separations between 
exact quantum query complexity and all classical complexity measures ($Q_E(f)=\tilde{O}(\sqrt{D(f)})$, 
$Q_E(f)=\tilde{O}(\sqrt{R_0(f)})$, and $Q_E(f)=\tilde{O}(R_2^{2/3}(f))$), between bounded-error quantum and zero-error
probabilistic complexity ($Q_2(f)=\tilde{O}(\sqrt[3]{R_0(f)})$), and between polynomial degree and randomized query complexity
($\adeg(f)=\tilde{O}(\sqrt[4]{R_2(f)})$) \cite{A+16}.

\section{Total functions: cheat sheet method}

\subsection{Query complexity}
\label{sec:cs}

After the developments described in the previous section, the biggest separation between quantum and randomized complexities still remained $Q(f)=O(\sqrt{R_2(f)})$. This was  improved to $Q(f)=\tilde{O}(R_2(f)^{2/5})$ in a breakthrough paper by Aaronson, Ben-David, and Kothari \cite{ABK}, using another new method, {\em cheat sheets}.

The key feature of {\em cheat sheet} method is that it takes separations for partial functions and transforms them into separations for total functions, by adding extra variables that allow to check that the input satisfies the promise for one of two cases when
the partial function $f$ is defined. The main result is

%For example, applying the cheat sheet construction to the partial function $f$ with $Q(f)=1$ and
%$R(f)=\Omega(\sqrt{N}/\log N)$ (described in section \ref{sec:partial}) yields $Q(f)=\tilde{O}(R_2(f)^{2/5})$. Moreover, resolving
%Problem \ref{prob:gap} in affirmative would immediately improve this separation to $Q(f)=\tilde{O}(R_2(f)^{1/3})$.

\begin{theorem}
\cite{ABK}
\label{thm:abk}
Let $f(x_1, \ldots, x_N)$ be a partial function with $Q_2(f)=Q$, $R_2(f)=R$ and $C(f)=C$. 
Then, there exists a total function $f_{CS}$ with $Q_2(f_{CS})=\tilde{O}(Q + \sqrt{C})$ and $R_2(f_{CS})=\Omega(R)$. 
\end{theorem}

Let $f(x_1, \ldots, x_{N^3})$ be the partial function $f=AND \circ OR \circ FORRELATION$ obtained by composing $AND$, $OR$ and $FORRELATION$ on $N$ variables each. 
From the complexities of $AND$, $OR$ and $FORRELATION$ and composition properties of the complexity measures it follows that $Q_2(f)=O(N)$, $C(f)=O(N^2)$ and $R_2(f)=\tilde{\Omega}(N^{2.5})$, implying 

\begin{theorem}
\cite{ABK}
There exists a total Boolean function $f_{CS}$ with $Q_2(f_{CS})=\tilde{O}(R_2^{2/5}(f_{CS}))$.
\end{theorem}

Moreover, if Problem \ref{prob:gap} was resolved in affirmative, we could substitute the corresponding $g$ instead of FORRELATION and Theorem \ref{thm:abk} would imply $Q_2(g_{CS})=\tilde{O}(R_2^{1/3+o(1)}(g_{CS}))$.

The cheat sheet method also gives new separations betwen $Q_2(f)$ and many of combinatorial complexity measures: 
$Q_2(f) = \tilde{\Omega} (C^2 (f))$,
$Q_2(f) = \tilde{\Omega} (\deg^2 (f))$, and 
$Q_2(f) = \Omega ({\adeg}^{4-o(1)} (f))$.
Moreover, several of results proven via the pointer function method (for example,  the $Q_E(f)=\tilde{O}(R_2^{2/3}(f))$ separation) can be reproven via cheat sheets \cite{ABK}.

To show Theorem \ref{thm:abk}, Aaronson et al. \cite{ABK} define the {\em cheat-sheet} function $f_{CS}$ in a following way. $f_{CS}$ has $tN+2^t t C \lceil \log N +1\rceil$ variables 
(for an appropriately chosen $t=\Theta(\log N)$) which we denote $x_{11}, \ldots, x_{tN}$, $y_{11}, \ldots, y_{2^t M}$
(where $M=C t \lceil \log N +1\rceil$). We interpret the blocks of variables
$x^{(i)}=(x_{i1}, \ldots, x_{iN})$ as inputs to the function $f$ and and the blocks $y^{(i)}=(y_{i1}, \ldots, y_{iM})$
as descriptions for $t$ certificates of function $f$, with the description containing both the set of variables $S\subseteq [N]$ and
and the values that $x_i$, $i\in S$ must take. (We refer to those blocks as {\em cheat-sheets}, as they allow to verify
the values of $f(x^{(1)})$, $\ldots$, $f(x^{(t)})$ with less queries than it takes to compute them.)

We interpret the $t$-bit string $s= s_1\ldots s_t$, $s_i=f(x^{(i)})$ as an index for the block $y^{(s)}$. 
We define that $f_{CS}=1$ if the block $y^{(s)}$ contains certificates for $f(x)=s_1$, $\ldots$,  $f(x)=s_t$
and the values of corresponding input variables in inputs $x^{(1)}, \ldots, x^{(t)}$ match the ones specified by the corresponding certificate.
Otherwise, $f_{CS}=0$.

To compute $f_{CS}$ by a quantum algorithm, we proceed as follows:
\begin{enumerate}
\item
compute $f(x^{(1)})$, $\ldots$, $f(x^{(t)})$, repeating each computation $O(\log t)$ times, to make the error probability at most $1/(10 t)$ for each $f(x^{i})$ (then, the probability that all $f(x^{(i)})$ are all simultaneously correct is at least 9/10);
\item
check whether the certificates in the block $y^{(s)}$ are satisfied by inputs $x^{(1)}, \ldots, x^{(t)}$, by using Grover's quantum search to search for a variable in one of $x^{(i)}$ which does not match the corresponding certificate.
\end{enumerate}
The complexity of the $1^{\rm st}$ stage is $O(Q t \log t)$. The complexity of the $2^{\rm nd}$ stage is $O(\sqrt{Ct} \log N)$, since we have to search among $tC$ variables $x^{(i)}_j$ ($t$ certificates, each of which contains $C$ variables),
Grover's quantum search \cite{Grover} allows to search among them by testing $O(\sqrt{tC})$ possibilities, and testing each possibility requires reading $O(\log N)$ variables in the block $y^{(s)}$. %If $Q$ is small, the complexity is dominated by the second stage. 
Thus, the overall complexity is $\tilde{O}(Q+\sqrt{C})$ quantum queries.

Classically, $Rt$ queries are required to solve $t$ instances of $f(x^{(i)})$. Moreover, if the number of queries is substantially smaller (of an order $o(Rt)$), then, with a high probability, most of $f(x^{(i)})$ are not solved yet and, at that point, a
classical algorithm cannot make use of certificate descriptions in $y^{(j)}$ because there are too many possible  $y^{j)}$. 
This suggests that $R_2(f_{CS})=\Omega(Rt)$ and Aaronson et al. \cite{ABK} show that this is indeed the case.

\comment{
Let $f(x_1, \ldots, x_{N^3})$ be the partial function $f=AND \circ OR \circ FORRELATION$ obtained by composing $AND$, $OR$ and $FORRELATION$ on $N$ variables. 
Then, we have
\begin{enumerate}
\item
$Q_2(f)=O(N)$, since $Q_2(AND)=Q_2(OR)=O(\sqrt{N})$ and $Q_2(FORRELATION)=1$;
\item
$C(f)=O(N^2)$, since $C(AND\circ OR)=N$ and $C(FORRELATION)\leq N$;
\item
$R_2(f)=\Omega(N^{2.5})$ (AND and OR require $\Omega(N)$ queries each, FORRELATION requires $\Omega(\sqrt{N})$ queries and it can be shown that no randomized algorithm for $f$ 
is more efficient than the compsition of randomized algorithms for AND, OR and FORRELATION).
\end{enumerate}
Together with the results described above, this yields $Q(f_{CS})=O(N)$ and $R_2(f_{CS})=\Omega(N^{2.5})$. 
If  Problem  \ref{prob:gap} is resolved positively, we will be able to substitute the corresponding function instead of FORRELATION and
obtain $f_{CS}$ with  $Q(f_{CS})=\tilde{O}(N)$ and $R_2(f_{CS})=\Omega(N^{3-o(1)})$. }

%We note that the idea of transforming partial function separations into total
%function separations was already present in \cite{ABB+} which 
%obtained $R_2(f)=\tilde{O}(R_0^{1/2}(f))$ by starting with a partial function for which $R_2(f)$ is asymptotically smaller than $R_0(f)$ and using pointers to transform it into a total function. The cheat sheet method, however, takes advantage of
%this idea in much more systematic way.

\subsection{Communication complexity}

The cheat sheet method has also found applications in a different domain, {\em communication complexity} \cite{KN,LS}. 
In the standard model of communication complexity, we have two parties, Alice and Bob, who want to compute a function $f(x, y)$, with
Alice holding the input $x$ and Bob holding the input $y$. The task is to compute $f(x, y)$ with the minimum amount of communication between Alice and Bob.
Communication complexity has a number of applications, from designing efficient communication protocols for various tasks to proving lower bounds on other models of computation (for example, streaming algorithms).

If quantum communication is allowed, the communication complexity may decrease exponentially. Similarly to query complexity, let $Q_2(f)$, $Q_E(f)$ and $R_2(f)$ denote the bounded-error quantum, exact quantum and bounded error
randomized communication complexity of $f$. A partial function with an exponential gap between $R_2(f)$ and $Q_2(f)$ was first constructed by Raz in 1999 \cite{Raz}. In a later work, it was shown that quantum protocols can be exponentially more efficient
even if the quantum protocol is restricted to one message from Alice to Bob \cite{RK} but it is compared against randomized protocols that can send an arbitrary number of messages back and forth.

However, similarly to query complexity, quantum advantages for total functions have been much more limited, with the best known separation of $Q(f)=O(\sqrt{R_2(f)})$ \cite{BCW,AA03} for the set disjointness problem 
which is the natural communication counterpart of Grover's search.
Anshu et al. \cite{Anshu} have adapted the cheat sheet method to communication complexity, proving

\begin{theorem}
\cite{Anshu}
\begin{enumerate}
\item
There is a total function $f(x, y)$ with $Q_2(f)=\tilde{O}(R_2^{2/5}(f))$;
\item
There is a total function $f(x, y)$ with $Q_E(f)=\tilde{O}(R_2^{2/3}(f))$;
\end{enumerate}
\end{theorem}

\section{Quantum-classical separations on almost all inputs?}

All known  partial functions $f(x_1, \ldots, x_N)$ with a superpolynomial quantum advantage have the property that $f$ takes one of values $f=0$ and $f=1$ on a very small subset of inputs. 
For example, for FORRELATION, the fraction of inputs with $f=1$ is exponentially small in the number of variables $N$. %The same is true for Simon's problem, period finding and all the other examples 
%where superpolynomial speedups are known. 
This had led to a following conjecture (known as a folklore since about 1999): 

\begin{conjecture}
\label{conj:aa}
\cite{AA14}
Let $\Q$ be a quantum algorithm that makes $T$ queries and let $\epsilon, \delta>0$. There is a deterministic algorithm with a number of queries that is polynomial in $T$, $\frac{1}{\epsilon}$ and $\frac{1}{\delta}$ and 
approximates the probability of $\Q$ outputting 1 to an additive error $\epsilon$ on at least $1-\delta$ fraction of all inputs.
\end{conjecture}

For total function, this conjecture implies that quantum and deterministic complexity are polynomially equivalent in the setting of approximately computing $f$. 
That is, for a total function  $f$, let $D_{\epsilon}(f)$ and $Q_{\epsilon}(f)$ be the smallest number of queries for a (deterministic or quantum) algorithm that outputs the correct answer on at least $1-\epsilon$ fraction of inputs $(x_1, \ldots, x_N)$.
Then, Conjecture \ref{conj:aa} implies that $D_{\epsilon}(f)$ and $Q_{\epsilon'}(f)$ are polynomially related, for all constant $\epsilon, \epsilon'$ with $\epsilon>\epsilon'$. 

There is a natural path towards proving Conjecture \ref{conj:aa}. Due to Lemma \ref{lem:poly}, Conjecture \ref{conj:aa} is implied by

\begin{conjecture}
\label{conj:aa1}
\cite{AA14}
Let $p(x_1, \ldots, x_N)$ be a polynomial of degree $2T$ which satisfies $|p(x_1, \ldots, x_N)|\leq 1$ for all $x_1, \ldots, x_N\in\{0, 1\}$
and let $\epsilon, \delta>0$. 
There is a deterministic algorithm with a number of queries that is polynomial in $T$, $\frac{1}{\epsilon}$ and $\frac{1}{\delta}$ and 
approximates $p(x_1, \ldots, x_N)$ to an additive error $\epsilon$ on at least $1-\delta$ fraction of all inputs.
\end{conjecture}

The natural way to design such a deterministic algorithm is by repeatedly choosing the variable $x_i$ that has the biggest influence on the value of $p$
(with the influence defined as $Inf_i(p)= E_x[|p(x)-p(x^{(\{i\})})|^2]$ with the expectation over a random choice of $x\in\{0, 1\}^n$).
To prove Conjecture \ref{conj:aa1}, it suffices to show 

\begin{conjecture}
\label{conj:aa2}
\cite{AA14}
Let $p(x_1, \ldots, x_N)$ be a polynomial of degree $2T$ which satisfies $|p(x_1, \ldots, x_N)|\leq 1$ for all $x_1, \ldots, x_N\in\{0, 1\}$. Assume that
\[ E_{x\in\{0, 1\}^n} \left[ \left( p(x) - E[p(x)]\right)^2 \right] \geq \epsilon .\]
Then, there is a variable $i$ with $Inf_i[p] \geq \left( \frac{\epsilon}{T}\right)^c$ for some constant $c$.
\end{conjecture}

Conjecture \ref{conj:aa2} connects with research in the analysis of Boolean functions. In particular, work of Dinur et al. \cite{DFKO} implies
a weaker form of the conjecture, with $Inf_i[p] \geq \frac{\epsilon^3}{2^{O(T)}}$. Improving it to  $Inf_i[p] \geq \left( \frac{\epsilon}{T}\right)^c $ is a 
challenging open problem which is interesting for both analysis of Boolean functions and quantum query complexity.

\section{Structure of quantum speedups?}

Another related question is: when can we achieve large  quantum speedups? 
%As we saw in sections \ref{sec:partial} and \ref{sec:total}, this is possible for
%partial functions but not for total functions. Also,  f
From the known examples of 
exponential and superexponential speedups for partial functions, we can observe that they are typically achieved for
problems with an algebraic structure. For example, Simon \cite{Simon} showed an exponential speedup
for the following problem:

{\bf Simon's problem.} Let $N=2^n$. We are promised that the input $(x_0, \ldots, x_{N-1})$ (where $x_i\in [M]$) 
satisfies one of two promises:
\begin{enumerate}
\item[(a)]
the mapping $i\rightarrow x_i$ is 2-to-1 with some 
$z\in[N], z\neq 0$ such that $x_y=x_{y\oplus z}$ for all $y\in [N]$, with $\oplus$ denoting bitwise addition 
modulo 2;
\item[(b)] 
the mapping $i\rightarrow x_i$ is 1-1.
\end{enumerate}

As shown by Simon, $Q_2(f)=O(n)$ but $R_2(f)=\Omega(2^{n/2})$. However, randomly permuting inputs turns Simon's problem into the problem of distinguishing whether $i\rightarrow x_i$ is 2-1 or 1-1 for which it is known that $Q_2(f)=\Theta(2^{n/3})$ but $R_2(f)=\Theta(2^{n/2})$ \cite{BHT,AA}, with the exponential quantum speedup disappearing. Similarly, permuting the input variables destroys the superexponential quantum speedup for the FORRELATION problem.

This leads to a question: can we show that quantum speedup is at most polynomial for any partial function that is symmetric with respect to permuting the input variables $x_i$? A positive answer would imply that large quantum speedups require problems with a structure (typically, of algebraic nature) that disappears if inputs are permuted. 

For the case when $x_i$'s are binary, evaluating a partial symmetric function essentially requires counting the number of $i:x_i=1$ up to a certain precision (which is sufficient for distinguishing whether the input $x=(x_1, \ldots, x_N)$ satisfies $f(x)=0$ or  $f(x)=1$). Quantum algorithms can count $i:x_i=1$ quadratically faster than classical algorithms \cite{Counting} and it is easy to show that
larger speedups cannot be obtained.

For non-binary inputs there are two possible ways of defining a ``symmetric function":
\begin{enumerate}
\item[(a)]
$f: [M]^N \rightarrow \{0, 1\}$ is symmetric,  if $f(x_1, \ldots, x_N) = f(x_{\pi(1)}, \ldots, x_{\pi(N)})$ for any permutation $\pi$ on $\{1, 2, \ldots, N\}$;
\item[(b)]
$f: [M]^N \rightarrow \{0, 1\}$ is symmetric,  if $f(x_1, \ldots, x_N) = f(\tau(x_{\pi(1)}), \ldots, \tau(x_{\pi(N)}))$ 
for any permutations $\pi$ on $\{1, 2, \ldots, N\}$ and $\tau$ on $\{1, 2, \ldots, M\}$.
\end{enumerate}

For example, the property of being 1-1 or 2-1 is preserved both if $x_1, \ldots, x_N$ are permuted and if the values for $x_1, \ldots, x_N$ are permuted. Thus, it is symmetric in the second, stronger sense. Similarly, element distinctness (determining whether $x_1, \ldots, x_N$ are all distinct) and other natural properties are symmetric in the second sense. For such properties, we have

\begin{theorem}
Assume that a partial function $f: [M]^N \rightarrow \{0, 1\}$ is symmetric in the second sense. Then, $R_2(f) = O(Q_2^7(f) \log^c Q_2(f))$.
\end{theorem}

It has been conjectured since about 2000 that a similar result also holds for $f$ with a symmetry of the first type.

A related question has been studied by Aaronson and Ben-David \cite{AD}: given a total function $f:\{0, 1\}^N \rightarrow \{0, 1\}$, 
can we define a subproblem $f_P$ ($f$ restricted to some subset $P \subseteq \{0, 1\}^N$) for which $Q_2(f_P) = O(\log^c R_2(f_P))$? 

For example, if $f(x_1, \ldots, x_N)=x_1 OR \ldots OR x_N$, then, for any restriction, the quantum advantage is at most quadratic.
(An intuitive explanation is that computing OR is essentially equivalent to finding $i:x_i=1$ and, for search, the quantum advantage is
quadratic whatever the number of $i:x_i=1$ is.)  In contrast, both MAJORITY and PARITY can be restricted so that quantum advantage becomes exponential.

The next theorem gives a full characterization when superpolynomial speedups can be achieved:

\begin{theorem}
\cite{AA14}
A promise $P \subseteq \{0, 1\}^N$ with $Q_2(f_P) = O(N^{o(1)})$ and $R_2(f_P)=\Omega(N^{\Omega(1)})$ exists
if and only if, for some $c>0$, there are $2^{N^c}$ inputs $x\in\{0, 1\}^N$ with $C_x(f)\geq N^c$.
\end{theorem}

\section{From polynomials to quantum algorithms}

As shown by Lemma \ref{lem:poly}, a quantum algorithm that makes $k$ queries can be converted into a polynomial of degree 
at most $2k$. %This observation has been useful for proving quantum lower bounds on specific functions (by showing that a polynomial with the desired properties does not exist), for example, for the collision problem \cite{AS}.
In the opposite direction, the existence of a polynomial of degree $2k$ does not imply the existence of a quantum algorithm that makes $k$ queries.
As mentioned in section \ref{sec:cs}, there is a total $f$ with $Q_2(f)=\Omega(\adeg^{4-o(1)}(f))$ \cite{ABK}.

However, there is an interesting particular case in which polynomials and quantum algorithms are equivalent.

\begin{theorem}
\label{thm:aa}
\cite{AA+}
Let $f(x_1, \ldots, x_N)$ be a partial Boolean function.
Assume that there is a polynomial $p(x_1, \ldots, x_N)$ of degree 2 with the following properties:
\begin{itemize}
\item
for any $x_1, \ldots, x_N\in\{0, 1\}$, $0\leq p(x_1, \ldots, x_N) \leq 1$;
\item
if $f(x_1, \ldots, x_N)=1$, $p(x_1, \ldots, x_N) \geq \frac{1}{2}+\delta$; 
\item
if $f(x_1, \ldots, x_N)=0$, $p(x_1, \ldots, x_N) \leq \frac{1}{2}-\delta$. 
\end{itemize}
Then, $f(x_1, \ldots, x_N)$ can be computed by a 1-query quantum algorithm with the probability of correct answer at least $\frac{1}{2}+\frac{\delta}{3(2K+1)}$ where $K$ is the Groethendieck's constant \cite{Pisier}
(for which it is known that $1.5707... \leq K \leq 1.7822...$ \cite{Braverman}).
\end{theorem} 

The main ideas for the transformation from a polynomial to a quantum algorithm are as follows:
\begin{enumerate}
\item
For technical convenience, we assume that $x_i$ are $\{-1, 1\}$-valued (instead of $\{0, 1\}$-valued).
We start by transforming a polynomial $p(x_1, \ldots, x_N)$ into another polynomial 
\[ q(x_1, \ldots, x_N, y_1, \ldots, y_N) = \sum_{i, j} a_{i, j} x_i y_j \]
which satisfies $q(x_1, \ldots, x_N, x_1, \ldots, x_N) = p(x_1, \ldots, x_N)$ for all $x_1, \ldots, x_N\in\{-1, 1\}$
and $|q(x_1, \ldots, x_N, y_1, \ldots, y_N)|\leq 1$ for all $x_1, \ldots, x_N, y_1, \ldots, y_N \in\{-1, 1\}$.
\item
If the spectral norm $\|A\|$ of the matrix $A$ is small, then the polynomial $q$ can be transformed into a quantum algorithm:
\begin{lemma}
	\label{cl:alg}
	Let $A=(a_{ij})_{i\in [N], j\in[M]}$ with $\sqrt{NM}\|A\| \leq C$ and let
	\[
	q(x_1, \ldots, x_N, y_1, \ldots, y_M) = \sum_{i=1}^{N} \sum_{ j=1}^M a_{ij} x_i y_j .
	 \]
Then, there is
%
%	If a polynomial
%	\[
%	p(x_1, \ldots, x_n, y_1, \ldots, y_m) = \sum_{i=1}^{n} \sum_{ j=1}^m a_{ij} x_i y_j
%	\]
%	satisfies $p(x,y)\in [-1, 1]$ for all   $ x \in  \BCF^n $, $ y \in \BCF^m $, then for every $ \delta>0 $  there is
a quantum algorithm that makes 1 query to $x_1, \ldots, x_N$,  $y_1, \ldots, y_{M}$ and outputs 1 with probability
	\[
	r = \frac{1}{2} \left(1 + \frac{q(x_1, \ldots, x_N, y_1, \ldots, y_{M})}{C}
		\right) .
	\]
\end{lemma}

The quantum algorithm consists of creating a combination of quantum states 
$\ket{\psi}=\sum_{i=1}^N \frac{x_i}{\sqrt{N}} \ket{i}$ and 
$\ket{\phi}=\sum_{j=1}^M \frac{y_i}{\sqrt{M}} \ket{j}$, applying $U=\sqrt{NM} \cdot A$ to $\ket{\phi}$ and then 
using the SWAP test \cite{Fingerprinting} to estimate the inner product of $\ket{\psi}$ and 
$U\ket{\phi}$ which happens to be equal to the desired quantity
$\sum_{i=1}^{N} \sum_{ j=1}^M a_{ij} x_i y_j$.  

If $U$ is  unitary, we can apply this procedure as described. If $\|U\|=C>1$, $U$ is not unitary and 
cannot be applied directly. Instead, we design and apply a unitary transformation that is equal to 
$\frac{1}{C} U$ on a certain subspace.
\item
For the general case, a corollary of Groethendieck's inequality \cite{Pisier,AA+,AB+} implies that, if $a_{ij}$ are such that
$|\sum_{i=1}^{N} \sum_{ j=1}^M a_{ij} x_i y_j | \leq 1$ for all choices of $x_i\in\{-1, 1\}$ and $y_j\in\{-1, 1\}$, there exist $\vec{u} = (u_i)_{i\in N}$ and
$\vec{v} = (v_j)_{j\in M}$ such that $\|\vec{u}\|=1$, $\|\vec{v}\|=1$, $a_{ij} = b_{ij} u_i v_j$ for all $i \in [N], j \in [M]$ and
$B=(b_{ij})_{i, j}$ satisfies $\|B\|\leq K$. 

Then, we can perform a similar algorithm with quantum states 
$\ket{\psi}=\sum_{i=1}^N  u_i x_i \ket{i}$ and 
$\ket{\phi}=\sum_{j=1}^M v_j y_j \ket{j}$.
\end{enumerate} 

Following this work, it was shown \cite{AB+} that quantum algorithms are equivalent to polynomial representations by 
polynomials of a particular type. Namely, the accepting
probability of a $t$ query quantum algorithm is equal to a completely bounded form of degree $2t$. For $t=1$, representatios of $f$
by a completely bounded forms are equivalent to representations by general polynomials (implying Theorem \ref{thm:aa}) but this does not hold for $t\geq 2$.

\end{document}